 \documentclass[nohyper,12pt,letterpaper]{JHEP}
 \usepackage{epsfig}
 
 \title{Space-like Singularities and Thermalization}
 \author{T.\,Banks\\
 Department of Physics and SCIPP\\
 University of California, Santa Cruz, CA 95064\\
 E-mail: \email{banks@scipp.ucsc.edu}\\
 {\it and}\\
 Department of Physics and NHETC, Rutgers University\\
 Piscataway, NJ 08540}

 \author{W.\,Fischler\\
 Department of Physics \\
 University of Texas, Austin, TX\\
 E-mail: \email{fischler@zippy.ph.utexas.edu}}

 \abstract{We conjecture that space-like singularities are simply regions in which all available
 degrees of freedom are excited, and the system cycles randomly through generic quantum states in
 its Hilbert space.   There is no simple geometric description of the interior of such a region,
 but if it is embedded in a semi-classical space-time an external observer sees it as a black hole.
 Big Bang and Crunch singularities, for which there is no such
 embedding, must be described in purely quantum terms.   We
 present several possible descriptions of such cosmologies.
 }
 \received{} \accepted{}
 \preprint{hep-th{0606260}\\\\SCIPP-06-05/RU-06-05/UTTG-10-06 \\}
 
\begin{document}
String Theory has provided a resolution of a variety of time-like
singularities\cite{tlsing}.  The case of space-like singularities,
relevant for the interior of black holes, and for cosmology, has
proven far less tractable. In this essay we want to suggest a
general physical picture of space-like singularities, which may
provide a basis for more rigorous understanding.

The basic claim we would like to make is that a generic future
space-like classical singularity, corresponds in the quantum theory
to a situation in which all of the available physical degrees of
freedom associated with the region of space-time in which the
singularity occurs, are maximally excited, and brought into a state
resembling thermal equilibrium.

Let us begin by examining the case of a black hole in
asymptotically flat or AdS space-time.   From the point of view of
an external observer, this is a thermal system, with finite
entropy.  The internal geometry of the black hole is singular, and
has the form
$$ds^2 = - {{dt^2}\over ({t\over R_S} - 1 + {t^2 \over R^2 } )} + dr^2 ({t\over R_S} - 1 + {t^2 \over R^2 }
) + r^2 d\Omega^2 .$$ Small fluctuation analysis leads to the
conclusion that this geometry is unstable near the singularity. We
will consider two ways of analyzing the non-linear evolution of the
instability.   In the first, we model a fluctuation by the behavior
of two mass points following geodesics of the unperturbed metric.
Define the center of mass energy, $\sqrt{s}$, of these two mass
points by the norm of the sum of the momentum of the first, with the
parallel transport of the momentum of the second, to the position of
the first.  The Schwarzschild radius corresponding to this energy
always exceeds the impact parameter of the two geodesics, as the
singularity is approached\footnote{Similar and complementary
analyses have been done recently by giddings\cite{giddings}.}. Thus,
the internal observer witnesses multiple black hole formation, which
it interprets as thermalization of the degrees of freedom that it
can observe.  It is not hard to imagine that, {\it from the interior
point of view}, all the excitations are swept up into black holes as
the singularity is approached.

The other analysis, which leads to similar conclusions, is that of
\cite{BKL}.  These authors showed that solutions of Einstein's
equations near a generic space-like singularity are chaotic.  The
general principles of statistical mechanics again lead us to
suppose that the entire interior is thermalized.  Indeed, a more
general argument that suggests this conclusion, is that with any
definition of energy far from the singularity, the time dependence
of the singular solutions should excite states of arbitrarily high
energy. Most systems we know of in quantum mechanics have a huge
degeneracy of high energy states, and no quantum numbers which
prevent the system from exploring this entire degenerate subspace,
starting from generic initial conditions with a large expectation
value of the energy.   Our basic hypothesis in this paper is that
any internal description of the states of a black hole will have
this property.

Indeed, for many future space-like singularities, like the interior
of a black hole formed through collapse, or the Coleman-De Luccia
Crunch for negative c.c. ,   the covariant entropy bound tells us
that observers which eventually encounter the singular part of
space-time, can only detect a finite entropy.   Let us interpret
this as a bound on the total number of states necessary to describe
the quantum mechanics of such an observer. Then the interior
dynamics is described by a quantum system with time dependent
Hamiltonian $H(t)$ which becomes singular at some $t$ we choose to
call $t = 0$. Let $N \gg 1$ be the number of states of the system.
Consider the Lie algebra generated by the collection of $H(t)$ at
different times. If the Hilbert space is in a reducible
representation of this algebra, then the system has conservation
laws.   We do not expect this to be true near a space-like
singularity.

If the algebra is $su(N)$, the time evolution operator $U(t)$ must
cycle randomly through $SU(N)$ as $t \rightarrow 0$, because the
$SU(N)$ group is compact. The state of the system thus explores the
entire Hilbert space as $t \rightarrow 0$.  If it is some proper
subalgebra of $su(N)$, introduce the coherent states
$$| \omega^a > = e^{ i \omega^a T_a} |\psi >, $$ where $|\psi > $ is
any state in the representation.   Let $H$ be the stability subgroup
of $|\psi $\footnote{$H$ is a proper subgroup of $G$, because the
representation is irreducible.} . The inequivalent values of the
parameters $\omega^a $ are coordinates on the compact coset manifold
$G/H$. For any choice of $| \psi >$ the coherent states are an
overcomplete basis. of the Hilbert space. Our singular time
evolution operator corresponds to a space-filling trajectory on the
compact manifold $G/H$.   So again, we conclude that the system
explores its entire Hilbert space as the singularity is approached.

In conventional statistical physics, which deals primarily with
systems which have an energy conservation law, thermal systems are
those which explore their entire Hilbert space, subject to the
constraint of fixed energy.   Time averages in a thermal system are
the same as averages over the micro-canonical ensemble.   Similarly,
quantum systems with a finite number of states and a singular time
dependent Hamiltonian of the type described above will have time
averages equivalent to ensemble averages over the maximally
uncertain density matrix.   We will continue to use the words
thermal and thermalization for such systems, even though they have
no energy conservation law.

Notice that quantum mechanics, and the assumption of a finite number
of states, regularizes singularities without removing them.   As the
singularity is approached, the rapidity with which the system cycles
through all of its states increases without bound.   Thus, if we
define time averages with some fixed scale, we will find that they
become time independent and equal to ensemble averages in the
maximally uncertain density matrix.   This is not a particularly
interesting system, but after time averaging it seems to give a
perfectly well defined limit.

In many situations, we can argue that the number of states of the
system is finite at a space-like singularity by using our refined
version of the covariant entropy bound\cite{bousso} in which the
area of causal diamonds bounds the actual number of states rather
than the entropy of some unspecified density matrix.   This is the
only version of the bound which would seem to make sense in a
general time dependent situation.

These heuristic arguments lead to a resolution of the black hole
information paradox.   If they are correct, then both internal and
external observations lead to the expectation of complete
thermalization of the degrees of freedom of the black hole.  The
heuristic picture of black holes formed by collisions inside the
black hole suggests that the internal observer does not really
experience the stretching and shrinking of the interior
Schwarzschild geometry.   Rather it experiences a chaotic excitation
of all of its degrees of freedom.   We believe that semi-classical
space-time pictures are not an appropriate description of the
interior, which is better described by the random quantum mechanics
of a finite system.

For large AdS black holes, this thermalization conjecture is enough
to make the internal and external observers' descriptions compatible
with each other.    The internal observer cannot survive long enough
to see a small black hole decay\footnote{This is a coordinate
dependent statement, since there exist {\it nice slice} coordinates
on which the local curvature is everywhere small, and much of the
Hawing radiation has already passed observers located large but
finite distances from the horizon.   However, as pointed out by some
of the authors of \cite{multi},  the internal time resolution on
such slices is super-Planckian and such an observer must use the
full non-local apparatus of quantum gravity to describe the
system.}, but it can see the thermalization which leads to the
decay.

\section{Thermalization in the Big Crunch}

Horowitz and Polchinski\cite{hp} were the first to study black
holes formed by excitations of a model singular universe: the null
orbifold.   Banks and Fischler  \cite{blacrunch} extended these
arguments to infinite flat FRW crunches, and suggested that in
this case the black hole formation evolved to a {\it dense black
hole fluid}, a state of matter/space-time with equation of state
$p = \rho$ and maximal entropy in any causal diamond.   These
arguments were heuristic, but suggestive.

A much more detailed picture of the relation between Big Crunch
space-times and thermalization, was achieved in recent work of
Hertog and Horowitz\cite{hh} (HH).  These authors analyzed the Big
Crunch that appears in the Lorentzian continuation of the Coleman
De Lucia\cite{cdl} (CDL) bubble describing the decay of an AdS
space. There are many peculiar aspects to this situation, not
least of which is that the decaying AdS space is supersymmetric
and one would have thought it was stable.  HH explain this by
noting that the CDL solution does not satisfy the normalizability
condition for solutions which are associated with states in the
Hilbert space of the stable supersymmetric AdS theory.   Rather,
it corresponds to a change in the Hamiltonian of the boundary
field theory.  The addition to the Hamiltonian is the integral of
a marginal triple trace operator which is unbounded from below.
The CDL solution, with no additional perturbations, does not
correspond to a sensible quantum theory.

However, HH speculate that one can stabilize the boundary field
theory by adding what are known technically as {\it dangerous
irrelevant operators}.   That is, the superficially renormalizable,
but unbounded from below, theory is obtained as the infra-red limit
of a stable ultraviolet theory.   If $\Phi$ is the dimension one,
single trace operator whose cube couples to the CDL solution, then
HH propose a boundary effective potential of the form
$$V(\Phi ) = R^{-2} \Phi^2 + a \Phi^3 + {b\over M} (\Phi^4 ) + \ldots .$$

The mass scale $M \gg R^{-1}$ is the scale of the stable UV theory
which provides the definition of this system.   It has a stable
vacuum with massive excitations of scale $M$, at a non-zero value
of $\Phi$.   The claim of HH is that this same theory, when
compactified on a sphere of radius $R$ has a metastable state,
with energy at some scale of order $M$, above the ground state,
whose low-lying excitations at energy $ \ll M$ (above the
meta-stable ground state), are described by the low lying
excitations of the superconformal field theory. The meta-stable
state decays, with a rate approximately given by the CDL
calculation, into a highly excited state of the massive field
theory.   Quite generically, we expect the energy produced in this
decay to thermalize.   From the boundary field theory point of
view, the endpoint of the decay is a thermal state, at a
temperature of order $M$, of the massive field theory.   Note that
this state has a finite entropy, of order $(MR)^3$, because the
boundary field theory lives on a sphere of radius $R$.

This seems to be a completely sensible and general description of
the meaning of an AdS decay into a Big Crunch, from the point of
view of boundary field theory.   The final state may not have a
good bulk space-time description, but it is a sensible state in a
well-defined boundary field theory.   From our present point of
view, what is important about this result is that the Big Crunch
represents a complete thermalization of the degrees of freedom
previously associated with the Ads space-time.

From the bulk point of view, there is one peculiarity of the
boundary description.   In the regime where the CDL calculation is
valid, the decay rate is exponentially small (as a function of the
AdS radius in Planck units).   Thus, before the decay occurs, an
observer in the meta-stable vacuum can bounce photons off the
boundary of AdS space and verify that it is living in an infinite
space-time.   However, most of the high energy black hole states
that one would associate with this infinite space-time, do not
exist.   When the black hole energy becomes of order the effective
potential barrier, black holes can decay in times of order $R^{-1}$.
Thus, the meaning of the phrase "asymptotically AdS space-time" is
ambiguous for this system.  Gentle exploration of the geometry by
test particles, over times short compared to the decay time, reveal
what appears to be a space-time described by this phrase,  but does
not guarantee the existence of the full set of CFT states that we
have come to associate with such a geometry.  Most of the high
energy states do not exist.   This is obvious from the boundary
field theory point of view, and could presumably be reproduced from
the bulk point of view, by computing the CDL transition rate for AdS
black holes as a function of the black hole mass.  Most of the true
high energy states of this system are those of the massive boundary
field theory and probably do not have any useful bulk space-time
description\footnote{General arguments suggest that the UV entropy
of the fixed point theory defining the massive boundary correlations
should be larger than that of the SUSY theory.   This would imply
that if it had an AdS dual, that dual would have smaller c.c. than
the SUSY theory.   However, it would then be hard to understand the
CDL instanton as a space-time process.   It seems more likely that,
like free field theory, this UV fixed point does not have an AdS
dual.}.

One may even wonder about the degree of universality of the HH
proposal for the UV completion of the field theory corresponding
to the CDL space-time.   General renormalization group arguments
would suggest that it is not very unique.  Many UV theories have
the same IR behavior.   Before coming to this conclusion, it would
be well to understand better the one peculiar feature of the HH
proposal from the boundary field theory point of view.  The HH
proposal is not the conventional story of RG flow from a UV to an
IR fixed point.   The low energy excitations whose behavior is
governed by the CFT dual to the original AdS space-time, are
excitations of a highly excited meta-stable state of the massive
theory.  Furthermore this state only exists when the theory is
compactified on a sphere.  If we study the massive boundary theory
in flat Euclidean space, it has a unique vacuum with a correlation
length of order $M^{-1}$.  RG flow goes from the UV fixed point
(possibly Gaussian) and a trivial fixed point.   One must ask
whether a generic UV theory which flows to the same IR behavior in
flat space will have a meta-stable state on the sphere with the
required properties.

The answer to this question is quite unclear. To answer it, it
would be good to have a soluble example of a boundary field theory
with such a compactification induced meta-stable conformal state.

The HH proposal is the most rigorous argument we know that the
correct description of (at least certain) Big Crunches is in terms
of thermalization of available degrees of freedom.

\section{The Big Bang}

The space-like singularities discussed above all occur in the
future.  They are described in terms of an organization of the
states of the theory in terms of some Hamiltonian, with the
assumption that the initial state has fairly low entropy. This
cannot be the description of the Big Bang singularity if we want a
theory which contains the concept of a particle horizon, and/or
explains how the universe began in a state of low entropy, in such
a way that the cosmological and thermodynamic arrows of time point
in the same direction.

The authors of \cite{holocosm}constructed a general formalism for
constructing quantum cosmologies consistent with the existence of a
particle horizon and the holographic entropy bound.  Then, using the
BKL results on chaotic behavior and the Problem of Time as
motivation, they argued that a given observer near the Big Bang
should be described by a random sequence of Hamiltonians. The
variables of the system are fermionic oscillators, which were
interpreted as pixels of the holographic screen of the observer's
causal diamond.   If the probability distribution from which the
Hamiltonian is chosen, is concentrated in the
neighborhood\footnote{Here neighborhood is understood in the sense
of the renormalization group.   The large fermion limit of random
quadratic fermion systems is the massless $1 + 1$ dimensional Dirac
equation.  For large $N$ the probability distribution must be
concentrated in the RG basin of attraction of this fixed point
theory.} of Hamiltonians quadratic in the fermions, then the system
obeyed all the rules of quantum cosmology, and exhibited the scaling
laws of the flat FRW universe with equation of state $p=\rho$.

The $p=\rho$ universe is thus a generic initial state for a quantum
system obeying the rules of \cite{holocosm}.   In this sense it is a
generalization of the completely thermalized states we encountered
in our description of future space-like singularities.   Only those
degrees of freedom within the particle horizon are thermalized at
any given time. There cannot be any real observers in a $p=\rho$
universe\footnote{We use the word observer in a very general sense.
It is a quantum system with many semiclassical observables (like a
large volume , cutoff, quantum field theory), which can be isolated
from the rest of the universe, and interact with it in a controlled
manner. Observers need have neither gender, nor consciousness.},
because all of the degrees of freedom describing any causal diamond
are collapsed into a large black hole. The idea of holographic
cosmology is that one should choose the initial state in as generic
a way as possible, consistent with the existence of real observers.
The earlier papers of \cite{holocosm} give a heuristic picture of
what this maximally entropic but {\it observaphilic} universe looks
like, designed to convince the reader that it resembles what we see
around us. In particular, the flatness, homogeneity and horizon
problems are explained in terms of properties of the $p=\rho$
universe, and the requirement that the observaphilic portion of the
universe be stable against ``decay" back to the more entropic
$p=\rho$ state. In order to explain the existence of CMB
correlations on our current horizon scale, as well as the fact that
fluctuations of all scales begin their sub-horizon oscillations with
zero velocity\cite{dodelson}, one must invoke a period of inflation.

Holographic cosmology provides the only known clues to the question
of why the observable part of the universe began in a low entropy
initial state.   It identifies the generic initial condition as one
in which all parts of the universe are in continuous interaction,
and no isolated observers are possible. The initial conditions for
the observable universe are supposed to be those of maximal entropy,
consistent with the existence of observers.   The authors of
\cite{holocosm} claim that this implies a flat FRW universe, whose
energy density is initially dominated by an almost uniform
distribution of black holes. Given these initial conditions, and the
existence of scalar fields dynamically capable of giving slow roll
inflation, inflation is a reasonably probable outcome. There is no
other comparably robust explanation of why inflation began.

In particular, the oft-repeated statement that inflation only
needs special initial conditions in a small patch of the universe,
which then expands to become all we see, begs the question.  In
quantum field theory, the expanded patch contains many more
degrees of freedom than the original one.  At the initial time,
these degrees of freedom are not well described by effective field
theory, so one cannot make a reliable estimate of the probability
of inflationary initial conditions without a better understanding
of quantum gravity\footnote{In situations more controlled than the
early Universe, one knows that the attempt to create inflation in
a small patch, leads instead to the formation of black holes, with
microscopic entropy.}. The claim of holographic cosmology is that
inflation is not generic, but that the most generic observaphilic
initial conditions (which already explain the homogeneity and
flatness problems, without inflation) also predict a high
probability for inflation to begin if the dynamical equations
permit it.

From the point of view of the present paper, holographic cosmology
gives a prescription for past space like singularities which similar
to, but different than our description of future space like
singularities.  In constructing holographic cosmology we insisted on
incorporating the concept of particle horizon, which allows degrees
of freedom to interact only after ``they have come into causal
contact".

\section{Cyclic Universe?}

Various attempts to build cosmologies from the dynamics of string
theory moduli\cite{pbbekcyc}, have reopened the question of
whether the universe could have gone through one or more cycles of
Big Crunch followed by Big Bang.   If the picture of space-like
singularities proposed in this paper is correct then the answer to
this question is: {\it probably not in any useful sense}.   The
Big Crunch would be described by a thermalization of all degrees
of freedom in the universe.   There could not be any smooth
transition to a low entropy FRW initial condition.

Rather, the only way one could imagine moving from the Big Crunch
to such a low entropy state is through a rare thermal fluctuation.
This assumes that the entropy of the Big Crunch state is finite.
The HH model described above gives a very concrete (though
unrealistic) example of how this works.   The Big Crunch leads to
a finite entropy thermal density matrix of excitations of the true
vacuum of the boundary field theory.   By the principle of
detailed balance, there is a small probability for this state to
tunnel thermally to the meta-stable AdS state, after which the
whole Big Crunch story can begin again. This does describe a sort
of cyclic universe, which spends most of its time in a state which
has no bulk space-time description, accompanied by rare and brief
sojourns in a smooth space-time.

There is no sense in which the state of the system during the
Crunch time gives useful information about initial conditions in
AdS space. By their nature, the AdS fluctuations are rare, and
therefore singular (in the colloquial rather than mathematical
sense). Indeed, the inverse tunneling event can equally well lead
to any normalizable low energy excited state of the AdS theory.
The energy will thermalize around the AdS vacuum, and the whole
system will eventually decay into a thermal state of the true
vacuum. Since the meta-stable AdS state is (in cases where
semi-classical calculations are valid), highly excited from the
point of view of the true vacuum, the finite entropy of the state
it decays into will be large.   In that case, the initial AdS
configuration ``the next time around" is effectively
unpredictable.   This is not the kind of cyclic universe that has
been hypothesized in recent string inspired models.

An interesting question that arises is whether our own universe
might be describable as a rare fluctuation of a thermal state with
some fixed Hamiltonian.   Hypotheses like this probably go back to
Boltzmann. This hypothesis has recently been explored in the context
of a hypothetical quantum theory of stable dS space\cite{dyson}. The
conclusion of that study seems quite general. If we hypothesize that
the origin of the universe as a fluctuation then we conclude that
the state just after the fluctuation had maximal entropy compatible
with the existence of observers. The authors of \cite{dyson} and
most people who have thought about this problem, claim that this
principle is in contradiction with observation.  For example,
increase the entropy in such a way that the current microwave
temperature is ten times its observed value. It is then unlikely,
but possible, for nuclei to have survived photo-dissociation, in
such a way that stars produced a galactic environment conducive to
our type of carbon based life.   The probability for this to occur
is of order $10^{-n}$ with some relatively small value of $n$, but
because we have increased the entropy of the initial conditions by a
factor of $10^4$, there are $e^{10^4} 10^{-n}$ more ways to find
life arising in the ensemble of such universes, than in our own. Our
universe does not look like a generic fluctuation, constrained only
by the existence of observers.

Although this argument sounds persuasive, we feel that the question
deserves more study.

\section{Conclusions}

The hypothesis that the meaning of space-like singularities is
maximal thermalization of the available degrees of freedom, helps to
resolve the black hole information puzzle. Under this hypothesis,
the exterior and interior descriptions of the black hole state
coincide for a time: both sorts of observer see a thermalization of
all interior degrees of freedom. At this point the internal observer
ceases to exist, and the ultimate fate of the black hole is a
concept that only applies to the external observer. Thermalization
is also a plausible conclusion to a Big Crunch, and the model of
\cite{hh} provides a very explicit quantum mechanical model of how
this can occur.

The correct description of the beginning of the universe is less
clear and there seem to be two plausible hypotheses: holographic
cosmology and the origin of the universe as a thermal fluctuation.
Both could lead to a universe like our own if they could be shown to
predict a high probability for a period of slow roll inflation.
Holographic cosmology does claim to make such a prediction, if the
degrees of freedom of the universe include a scalar field with
appropriate Lagrangian. It is likely that a generic fluctuation
model does not, but the fluctuation idea is very general and there
could be implementations of it, which evade the ``Boltzmann's Brain"
paradox of \cite{dyson} and describe a universe like our own.

The interpretation of future space-like singularities in terms of
thermalization does not fit well with models which envision a
smooth and predictive transition between Big Bang and Big Crunch.
These models require a much more controlled resolution of
space-like singularities.

\section{Acknowledgements} TB and WF would like to thank G. Horowitz for conversations
about this paper.   The work of TB was supported in part by the DOE
under grant DE-FG03-92ER40689.  The research of WF was supported by
NSF under grant PHY 0071512 and 0455649.




  %




\end{document}